\documentclass[preprint,aps,prb,floats,showpacs,superscriptaddress]{revtex4}
\usepackage{graphicx,epsfig}
\usepackage{times}
\usepackage{graphics,dcolumn,bm,float}
\usepackage{amssymb,amsmath,rotate,color}

\def\dblone{\hbox{$1\hskip -1.8pt\vrule depth 0pt height 1.6ex width 0.7pt \vrule depth 0pt height 0.3pt width 0.12em$}}
\begin{document}

\title{Atomic defect states in monolayers of MoS$_2$ and WS$_2$}
\author{Saboura Salehi}
\affiliation{Department of Physics, Payame Noor University, P.O.
Box 19395-3697 Tehran, Iran}
\author{Alireza Saffarzadeh}\email{asaffarz@sfu.ca}
\affiliation{Department of Physics, Payame Noor University, P.O.
Box 19395-3697 Tehran, Iran} \affiliation{Department of Physics,
Simon Fraser University, Burnaby, British Columbia, Canada V5A
1S6}
\date{\today}

\begin{abstract}
The influence of atomic vacancy defects at different
concentrations on electronic properties of MoS$_2$ and WS$_2$
monolayers is studied by means of Slater-Koster tight-binding
model with non-orthogonal $sp^3d^5$ orbitals and including the
spin-orbit coupling. The presence of vacancy defects induces
localized states in the bandgap of pristine MoS$_2$ and WS$_2$,
which have potential to modify the electronic structure of the
systems, depending on the type and concentration of the defects.
It is shown that although the contribution of metal (Mo or W) $d$
orbitals is dominant in the formation of midgap states, the
sulphur $p$ and $d$ orbitals have also considerable contribution
in the localized states, when metal defects are introduced. Our
results suggest that Mo and W defects can turn the monolayers into
p-type semiconductors, while the sulphur defects make the system a
n-type semiconductor, in agreement with \textit{ab initio} results
and experimental observations.
\end{abstract}

\maketitle

\section{Introduction}
Layered transition metal dichalcogenides (LTMDs) have attracted
intensive attentions in recent years due to their intrinsic
non-zero bandgap, which gives them a superior advantage over
graphene for use in nanoelectronic and optoelectronic applications
such as field-effect transistors and electroluminescent devices
\cite{Wang,Ataca}. This class of layered materials with chemical
composition of MX$_2$, where M and X correspond to the transition
metal and the chalcogen elements, respectively, crystallizes in a
hexagonal structure like graphene in which the M-atom layer is
covalently bonded and sandwiched between the two X-atom layers.
Among LTMD materials, MoS$_2$ and WS$_2$ monolayers with a direct
bandgap configuration have been extensively investigated because
of many intriguing physical and chemical properties
\cite{Radisavljevic,Braga,ZhenZhou}. These compounds can be
synthesized through various methods, such as mechanical
exfoliation \cite{Novoselov}, chemical vapor deposition
\cite{Zhang}, and intercalation techniques \cite{Eda}. In
addition, they have quite similar lattice constants which also
enable the synthesis of MoS$_2$-WS$_2$ heterostructures with
minimum interfacial defects \cite{Kang,Kosmider,Chen,Yoo,Gong}.

Point defects such as atomic vacancies may cause a large variation
in the electronic and optical properties of LTMDs. Vacancy
defects, which can be created by thermal annealing and $\alpha$
particles \cite{Tongay} or electron beam irradiation \cite{Zhou},
form localized trap states in the bandgap region, leading to light
emission at energies lower than the interband optical transition
energy \cite{Tongay}. On the other hand, the observed charge
mobility in single-layer MoS$_2$ is surprisingly low compared to
bulk sample \cite{Novoselov,Radisavljevic}, indicating that the
charge carrier scattering by structural defects, such as vacancies
and grain boundaries, may be a primary source for such a low
mobility \cite{Tongay,Enyashin,Asl}. Hong \textit{et al.}
\cite{Hong} have studied point defects and their concentrations
for several samples of MoS$_2$ by means of different preparation
methods. They found that the dominant type of point defects in
each sample is strongly dependent on the chosen sample preparation
method. Nevertheless, the sulphur vacancy is the predominant point
defect compared to Mo vacancy, regardless of the type of
preparation method. \cite{Hong}

The effects of point defects on the electronic structure of LTMDs
have also been theoretically studied by several groups using first
principles calculations
\cite{Enyashin,Asl,Ataca2,Ma,Komsa,Wei,Liu,Zhou2,BHuang} and
6-band tight-binding (TB) model \cite{Yuan}. Although, \textit{ab
initio} methods based on density functional theory (DFT) can
achieve a good degree of accuracy to describe the electronic
structure of pristine LTMD materials, they are limited in their
application by the presence of defects in the samples. For
instance, simulation of vacancy-doped MoS$_2$ and WS$_2$
monolayers with a random distribution of vacancies requires a very
large supercell in the calculations which is computationally
expensive for DFT methods. With TB approach which is a simpler and
less computationally demanding method, however, it is possible to
deal with such large systems. The use of large supercells within
TB model makes it also possible to eliminate the vacancy-vacancy
interactions from the calculations.

After the two-band ${\bf k}\cdot{\bf p}$ model describing the
conduction and valence bands around the two valleys (K and
K$^\prime$ points) in the hexagonal Brillouin zone of LTMD
\cite{Xiao}, several TB models in various approximations have been
proposed to reproduce the first-principles band structure of
pristine LTMD \cite{Rostami,Zahid,Cappelluti,Roldan,Liu}. Among
them, the TB model of Zahid \textit{et al.} \cite{Zahid},
including nonorthogonal $sp^3d^5$ orbitals of M and X atoms and
spin-orbit coupling, is able to accurately reproduce the
first-principles bands for a wide range of energies in the
Brillouin zone. The model considers nearest-neighbor Slater-Koster
hopping matrix elements of M-M, M-X, and X-X, which can be applied
to monolayers, bilayers and bulk MX$_2$ \cite{Zahid}.

In this work, based on the parameterized TB model of Zahid
\textit{et al.} \cite{Zahid}, we explore the influence of vacancy
defects on the electric properties of MoS$_2$ and WS$_2$
monolayers to see how the missing atoms at different
concentrations evolve the intrinsic bandgap and electronic states
of the monolayers. Since the model presents an accurate
description for the band structure of LTMDs, the application of
this model to defective MoS$_2$ and WS$_2$ provides a more
realistic understanding of the electronic states contributing to
the process of vacancy formation and the accurate location of
defect states within the bandgap. Moreover, the optimized
geometries of the monolayers obtained by \textit{ab initio}
calculations have demonstrated that atomic vacancies do not cause
a considerable geometry deformation and the neighboring atoms
around the vacancies do not show any visible displacement
\cite{Wei,Ataca}. Therefore, the defect-induced deformation is
ignored. We show that the vacancy defects mainly induce localized
states within the bandgap of pristine MoS$_2$ and WS$_2$, leading
to a shift of the Fermi level toward valance or conduction band,
depending on the type of vacancy. The rest of this paper is as
follows. In section II we introduce our model and formalism for
calculation of band structure and electronic states of the
defective monolayers. Numerical results and discussion for
electronic properties of MoS$_2$ and WS$_2$ with different types
and concentrations of vacancy defects are presented in Sec. III. A
brief conclusion is given in Sec. IV.
\begin{figure}
\centerline{\includegraphics[width=0.9\linewidth]{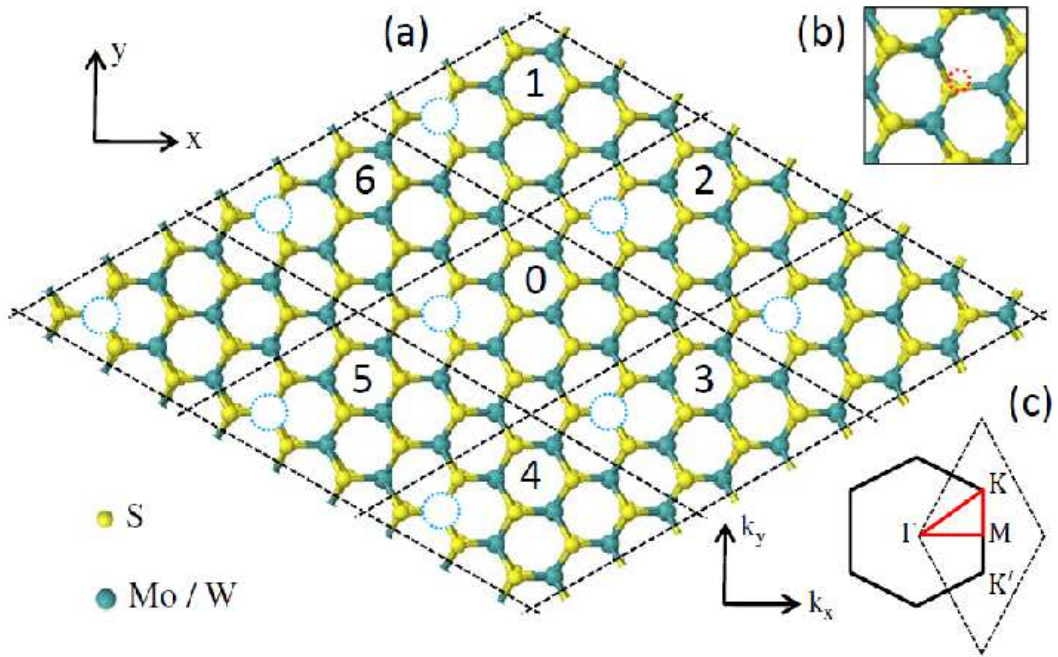}}
\caption{(Color online) (a) Top view of MoS$_2$ (WS$_2$) monolayer
with 3$\times$3 supercells containing Mo (W) vacancy (blue dotted
circles). Numbers 1-6 represent the supercells for which the
reference supercell, shown by $\mathbf{0}$, has any overlap. (b)
Top view of the region around a single S vacancy shown by red
dotted circle. (c) Hexagonal (solid) and rhombic (dotted)
Brillouin zones of the monolayer with the red lines along which
the band structures are calculated.}
\end{figure}

\section{Model and formalism}
The band structure of defective MoS$_2$ and WS$_2$ is carried out
within the non-orthogonal Slater-Koster scheme \cite{Slater}. From
DFT calculations \cite{Zahid}, we know that the bands of both
structures are made up of the $s$, $p$, and $d$ valence orbitals
of Mo, W, and S atoms. Therefore, a basis set consisting of $s$,
$p_x$, $p_y$, $p_z$, $d_{xy}$, $d_{xz}$, $d_{yz}$, $d_{x^2-y^2}$,
$d_{3z^2-r^2}$ orbitals is used as a starting point for
constructing the TB Hamiltonian. This means that for a monolayer
MX$_2$ with one M atom and two X atoms per unit cell, we should
consider a 27-band TB spinless model. Moreover, a Bloch sum is
taken into account for each atomic orbital on each atomic site in
the unit cell due to the periodicity of the monolayer. On the
other hand, in order to model a defective MX$_2$ monolayer with
different vacancy concentrations the system is partitioned into
$n\times n$ supercells each containing $n^2$ unit cells where $n$
is an integer number. Figs. 1(a) and 1(b) show such a monolayer
with 3$\times$3 supercells each containing a single vacancy
defect. Since all the valence orbitals of the atoms belonging to
the supercell are included in the atomic orbitals basis set, the
number of bands increases with the size of the supercell. The
total Hamiltonian of MX$_2$ monolayer can be written as
\begin{equation}\label{H}
\mathcal{H}=\mathcal{H}_{\mathrm{SK}}\otimes\dblone+\mathcal{H}_{\mathrm{SO}}\
,
\end{equation}
where $\mathcal{H}_{\mathrm{SK}}$ represents the Slater-Koster
tight-binding Hamiltonian for non-orthogonal $sp^3d^5$ orbitals,
$\dblone$ is the $2\times2$ identity matrix, and
$\mathcal{H}_{\mathrm{SO}}$ is an atomiclike spin-orbit coupling.
$\mathcal{H}_{\mathrm{SK}}$ has the same form for both spin-up and
spin-down states and can be expressed in the real space as
\begin{equation}\label{HSK}
\mathcal{H}_{\mathrm{SK}}=\sum_{i,j}\sum_{\alpha,\beta}(\epsilon_{i\alpha}\delta_{ij}
\delta_{\alpha\beta}+t_{i\alpha,j\beta})d^\dag_{i\alpha}d_{j\beta}\
,
\end{equation}
where $d^\dag_{i\alpha}$ is the creation operator for an electron
in an atomic valence orbital $\alpha$ at $i$-th atom,
$\psi_{i\alpha}$, with on-site energy $\epsilon_{i\alpha}$. The
hopping parameters,
$t_{i\alpha,j\beta}=\langle\psi_{i\alpha}|\mathcal{H}_{\mathrm{SK}}|\psi_{j\beta}\rangle$,
between atomic orbitals $\psi_{i\alpha}$ and $\psi_{j\beta}$ are
real Slater-Koster integrals that depend for each orbital pair on
the directional cosines of the vector connecting nearest neighbors
and on the Slater-Koster TB parameters $V_{ss\sigma}$,
$V_{sp\sigma}$, $V_{ps\sigma}$, $V_{pp\sigma}$, $V_{pp\pi}$,
$V_{sd\sigma}$, $V_{ds\sigma}$, $V_{pd\sigma}$, $V_{dp\sigma}$,
$V_{pd\pi}$, $V_{dp\pi}$, $V_{dd\sigma}$, $V_{dd\pi}$, and
$V_{dd\delta}$ for MoS$_2$ \cite{Zahid} and WS$_2$ \cite{Private}.
These parameters are related to hopping processes between
nearest-neighbor Mo-S (W-S), between the nearest-neighbor in-plane
Mo-Mo (W-W), and between the nearest-neighbor in-plane and
out-of-plane S-S atoms in MoS$_2$ (WS$_2$) monolayer. The hopping
terms between next nearest neighbors are ignored in this model.
\begin{figure}
\centerline{\includegraphics[width=0.6\linewidth]{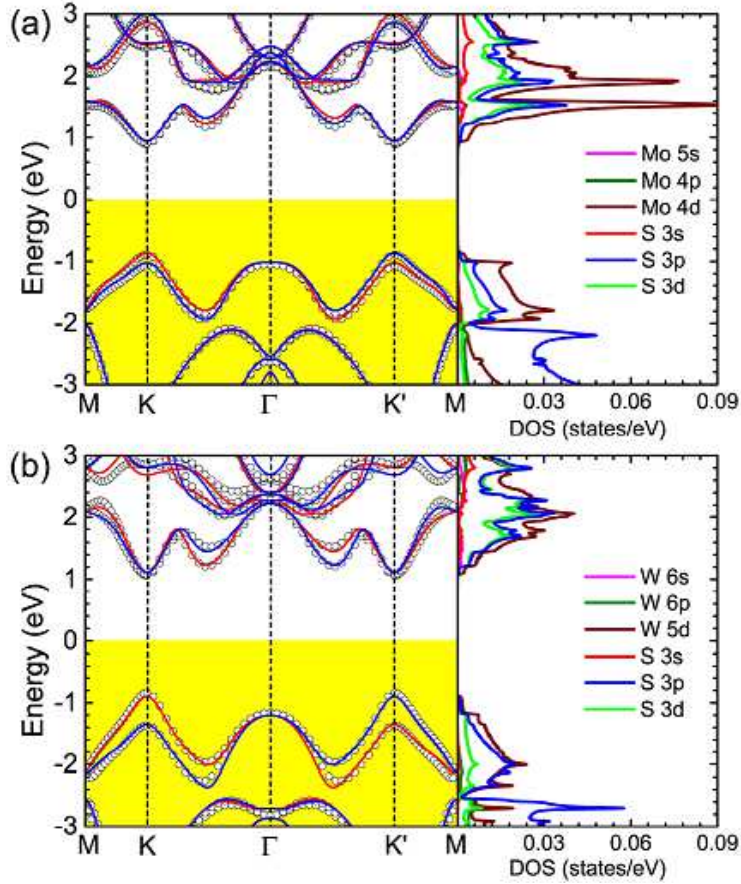}}
\caption{(Color online) (a) The calculated band structure with the
projection of spin operator and corresponding partial DOS of
pristine (a) MoS$_2$ and (b) WS$_2$ monolayers. The blue and red
colors in the band structure indicate the spin-up and spin-down
states, respectively. The hollow circles correspond to DFT
calculations \cite{Feng,Gibertini,Wickramaratne}. The intersection
of white and yellow regions shows the Fermi energy.}
\end{figure}

The intra-atomic spin-orbit interaction acting on both the
transition metal and the chalcogen atoms is incorporated in the
Hamiltonian via the second term in Eq. (\ref{H}) which is written
as \cite{Roldan},
\begin{equation}\label{HSO}
\mathcal{H}_\mathrm{SO}=\sum_{i}\sum_{\sigma\sigma'}\frac{\lambda_i}
{2\hbar}\mathbf{L}_{i}\cdot\bm{\tau}_{\sigma\sigma'}\ ,
\end{equation}
where $\bm{\tau}$ are the Pauli spin matrices, $\mathbf{L}_{i}$ is
the atomic angular momentum operator, and $\lambda_i$ is the
intra-atomic spin-orbit coupling constant which depends on the
type of atom $i$. In the presence of such a spin-orbit coupling,
inversion symmetry breaking in the LTMD materials lifts the spin
degeneracy of the energy bands, leading to a strong spin-splitting
in the valence-band maximum (VBM) \cite{Xiao}.

Within the non-orthogonal scheme, the orbital overlap
$\mathcal{S}_{i\alpha,j\beta}=\langle
\psi_{i\alpha}|\psi_{j\beta}\rangle$ obtained from Slater-Koster
parameters, can be non-zero. Therefore, the band structure of the
system is calculated by solving the generalized eigenvalue
problem:
\begin{equation}\label{HSK}
\mathcal{H}({\bf k})\mathcal{C}_\alpha({\bf
k})=E_\alpha\mathcal{S}({\bf k})\mathcal{C}_\alpha({\bf k})\  ,
\end{equation}
where $\mathcal{C}_\alpha({\bf k})$ denotes the eigenvector of the
band $\alpha$ and ${\bf k}$ is an allowed wave vector in the
two-dimensional Brillouin zone, shown in Fig. 1(c). Note that the
size of Hamiltonian $\mathcal{H}({\bf k})$ and overlap,
$\mathcal{S}({\bf k})$, matrices (including spin), which is the
same as the size of $\mathcal{H}$ and $\mathcal{S}$ in the real
space, is equal to $2N\times 2N$, where $N$ is the number of basis
orbitals per supercell and 2 is for spin. In the absence of
vacancy defects, the size of these matrices, including the
spin-orbit interaction for a monolayer with $3\times 3$
supercells, is $486\times 486$. To introduce a single vacancy
defect, we remove one atom from the supercell while the symmetry
of the lattice remains intact \cite{Papa}. This reduces the number
of atomic orbitals in each supercell and hence the size of
matrices.

The Hamiltonian and overlap matrices in the discrete version of
Bloch's theorem \cite{Datta} can be written as
\begin{equation}\label{HK}
\mathcal{H}({\bf k})=\sum_{m=0}^6\mathcal{H}_{0m}\mathrm{e}^{i{\bf
k}\cdot({\bf r}_m-{\bf r}_0)}\  ,
\end{equation}
\begin{equation}\label{HK}
\mathcal{S}({\bf k})=\sum_{m=0}^6\mathcal{S}_{0m}\mathrm{e}^{i{\bf
k}\cdot({\bf r}_m-{\bf r}_0)}\  ,
\end{equation}
where $m$ is the supercell index and the summation runs over all
neighboring supercells including the reference supercell, as shown
by $\mathbf{0}$ in Fig. 1(a). Because of the periodicity of the
lattice in Fig. 1(a), the result is independent of the reference
unit cell that we choose. The Green's function of MX$_2$ monolayer
is defined by
$G(\epsilon,\mathbf{k})=[(\epsilon+i\delta)\mathcal{S}(\mathbf{k})-\mathcal{H}(\mathbf{k})]^{-1}$
where $\delta$ is a positive infinitesimal. Accordingly, the local
density of states (DOS), $\rho^{\sigma}_{i\alpha}(\epsilon)$, for
an electron with spin $\sigma$ in an atomic orbital $\alpha$ at
site $i$ in the supercell can be obtained directly from the
Green's function of the MX$_2$ monolayer through \cite{Lopez}
\begin{equation}\label{rho}
\rho^{\sigma}_{i\alpha}(\epsilon)=-\frac{1}{\pi}\mathrm{Im}\sum_{\bf
k}[G(\epsilon,{\bf k})\mathcal{S}({\bf
k})]_{i\alpha,i\alpha}^{\sigma,\sigma}\  .
\end{equation}
Therefore, the partial DOS of an atomic orbital $\alpha$ in the
unit cell is simply given as
$\rho_{\alpha}(\epsilon)=\sum_{i,\sigma}\rho^{\sigma}_{i\alpha}(\epsilon)$.

\section{results and discussion}
We now use the method described above to study the influence of
single vacancy defects on electronic properties of MoS$_2$ and
WS$_2$ monolayers. First, we discuss the TB band structure and
partial DOS of the pristine monolayers in the presence of
spin-orbit interaction and their consistency with DFT
calculations. It should be mentioned that for more quantitative
agreement between our TB results and fully-relativistic ab-initio
DFT calculations \cite{Xiao}, we include the spin-orbit coupling
between Mo $d$ orbitals, instead of $p$ orbitals and only between
W $d$ orbitals, instead of $p$ and $d$ orbitals used in Ref.
\onlinecite{Zahid}. The fitted spin-orbit parameters, the
valence-band spin-splitting values, and the band gaps obtained in
this way are presented in Table I. The bandgap values of 1.80 eV
for MoS$_2$ and 1.98 eV for WS$_2$ and the valence band spin-orbit
splittings, obtained using this method, are in good agreement with
DFT \cite{Zahid,Feng,Kang} and experimental values \cite{Mak}.
Note that the single-layer WS$_2$ has a larger bandgap because the
crystal field splitting of the metal $d$ states, which is larger
in W compared to Mo, is responsible for a large part of the
bandgap \cite{Mattheiss}. In addition, the valence band
spin-splitting in WS$_2$ is almost three times larger than that in
MoS$_2$, which makes the observation of valley and spin Hall
effect easier in WS$_2$ \cite{Xiao}.

\begin{table}[h]
\caption{The fitted values of spin-orbit parameters,
$\lambda_{i}$, for Mo, W, and S atoms; the spin-splitting of the
VBM, $\Delta_{\mathrm{SO}}$; and the values of bandgap,
$E_\mathrm{g}$. All quantities are in units of eV.} \centering
\begin{tabular}{c c c c c}
\hline\hline &~~~~~$\lambda_{i,\mathrm{Mo/W}~~~~~}$ &
~~~~~$\lambda_{i,\mathrm{S}}$~~~~~ & ~~~~~
$\Delta_{\mathrm{SO}}$~~~~~ & ~~~$E_\mathrm{g}~~$\\[0.5ex]\hline
MoS$_2$ & 0.130 & 0.057 & 0.154 & 1.80 \\
WS$_2$  & 0.422 & 0.057 & 0.449 & 1.98 \\[1ex]
\hline\hline
\end{tabular}
\label{tab:1}
\end{table}

Figs. 2(a) and 2(b) show the electronic structure with the
projection of spin operator and the corresponding partial DOS of
MoS$_2$ and WS$_2$ monolayers, respectively. We see that both
monolayers have a direct bandgap at the two inequivalent corners K
and K$^\prime$ of the Brillouin zone (Fig. 1(c)). The spin
splittings in the band structures along $\Gamma$-K-M and
$\Gamma$-K$^\prime$-M lines are opposite which lead to
valley-selective optical absorption and may cause optically
induced valley and spin Hall effects \cite{Xiao}. Comparing the
band structures with that given in Refs. \onlinecite{Feng},
\onlinecite{Wickramaratne}, and \onlinecite{Gibertini} clearly
shows the quantitative agreement between TB and DFT results (see
hollow circles in Fig. 2). From the partial DOS of pristine
monolayers one can see the contribution of each type of atomic
orbitals to the formation of energy bands. The conduction-band
minimum (CBM) and the VBM of MoS$_2$ (WS$_2$) are mostly dominated
by Mo (W) $d$ orbitals and S $p$ orbitals, in agreement with DFT
\cite{Feng}. Moreover, the inclusion of S $3d$ orbitals in our
model leads to a nonzero contribution to the electronic states
which is comparable to S $3p$ orbitals in the conduction band.

To study atomic vacancy defects, the monolayer is partitioned into
supercells and one atom from each supercell is removed without any
change in the symmetry of the lattice (see Fig. 1(a)). We have
examined several supercell sizes (3$\times$3, 4$\times$4 and
5$\times$5) to reveal the strength of vacancy-vacancy interaction
on the localized midgap states. Note that in the supercell
calculations, as the supercell grows in size, the corresponding
Brillouin zone in the ${\bf k}$-space shrinks and the bands in the
original (normal) Brillouin zone get folded into the supercell
Brillouin zone. In other words, if the supercell is $n$ times
larger than the normal cell, the Brilouin zone of the supercell
will be $n$ times smaller and will contain $n$ times more bands.

Now, let us consider metal vacancies at different concentrations
in MoS$_2$ and WS$_2$ monolayers. In Fig. 3, we show the band
structure and the corresponding partial DOS of MoS$_2$ when a Mo
vacancy is introduced. Since each supercell contains only one
atomic vacancy defect, the size of supercell manifests itself as a
measure of defect concentration. Accordingly, the Mo defect
concentration per supercell in Fig. 3(a)-3(c) is $\frac{1}{27}$,
$\frac{1}{48}$, and $\frac{1}{75}$, respectively. The defect
concentrations represent the ratio of number of vacancies to the
number of atomic sites per supercell which correspond to vacancy
densities of $\sim 12.8\times 10^{13}$, $7.2\times 10^{13}$, and
$4.6\times 10^{13}$ cm$^{-2}$, respectively. We see that at high
vacancy concentration, i.e., $\frac{1}{27}$, the defect states
form a band in the middle of the gap, whose width is $\sim$ 0.73
eV (see Fig. 3(a)). The midgap band creates defect states with
three peaks in the DOS spectrum arising from neighboring Mo $4d$
orbitals and S $3p$ and $3d$ orbitals around the defect. In
addition, the vacancy defect induces a sharp peak at the top of
the valence band (Fig. 3(a)), corresponding to S $3p$ orbitals,
which shifts toward lower energies as the concentration decreases.
The midgap band splits into two bands centered around Fermi level
of pristine MoS$_2$ as shown in Fig. 3(b). These bands become more
localized at concentration $\frac{1}{75}$ (Fig. 3(c)) indicating
that gap states generated by Mo vacancies are mainly localized
around atomic defects, in agreement with Ref. \onlinecite{Yuan}.
It is important to point out that the position of Fermi energy is
determined by counting the number of electrons that the atoms in
the supercell provide. These electrons fill up the lowest energy
bands and hence, the Fermi level lies between the highest occupied
band and the lowest unoccupied band.
\begin{figure}
\center\includegraphics[width=1.0\linewidth]{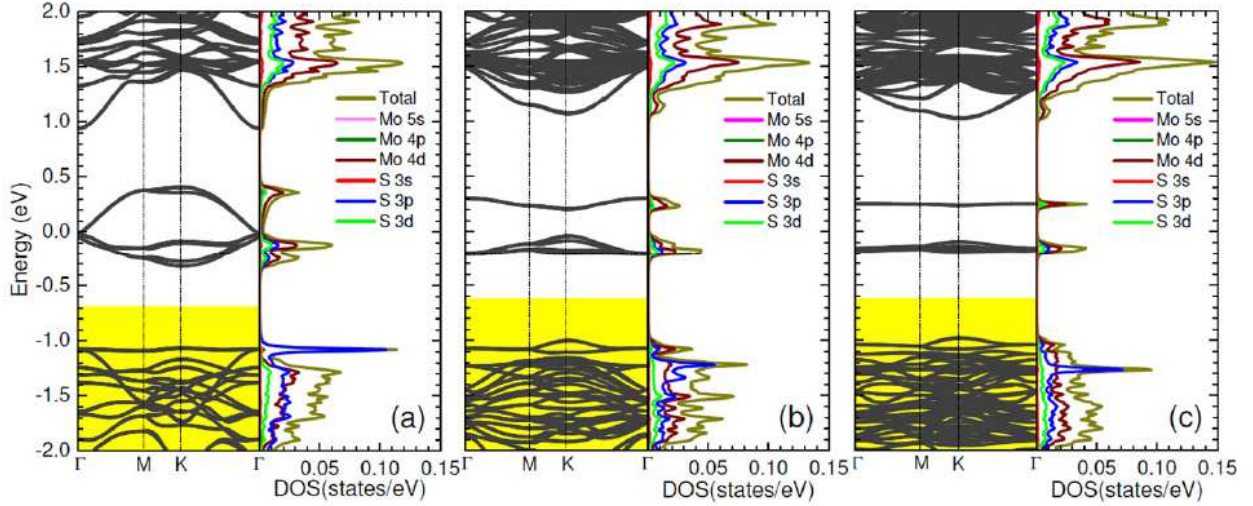}
\caption{Calculated band structure and corresponding partial DOS
of MoS$_2$ monolayer with (a) 3$\times$3, (b) 4$\times$4, and (c)
5$\times$5 supercells each containing a single Mo vacancy. The
intersection of white and yellow regions shows the Fermi energy.}
\end{figure}

We now consider the effect of metal vacancies on electronic
structure of WS$_2$ monolayer at different defect concentrations,
as shown in Figs. 4(a)-4(c). The band structure of Fig. 4(a) shows
two midgap bands in their close proximity, located below the Fermi
energy of the pristine monolayer. These bands form a single narrow
band with three peaks in the partial DOS, associated with
localized states around the defects as the concentration decreases
(see Fig. 4(b) and 4(c)). We see that the midgap states originate
mainly from W $5d$ orbitals and S $3p$ and $3d$ orbitals,
indicating that the contribution of S $3d$ orbitals could be
considerable in the electronic structure of LTMD when metal
vacancies are introduced in the system. Moreover, contrary to the
electronic structure of defective MoS$_2$ with Mo vacancies, the
induced sharp peak at the top of the valence band, corresponding
to the S $3p$ orbitals, is not shifted down in energy as the
distance between point defects in WS$_2$ monolayer increases. This
reveals a strong hybridization between sulphur atoms, and hence,
localization of S $3p$ states around W defects.
\begin{figure}
\center\includegraphics[width=1.0\linewidth]{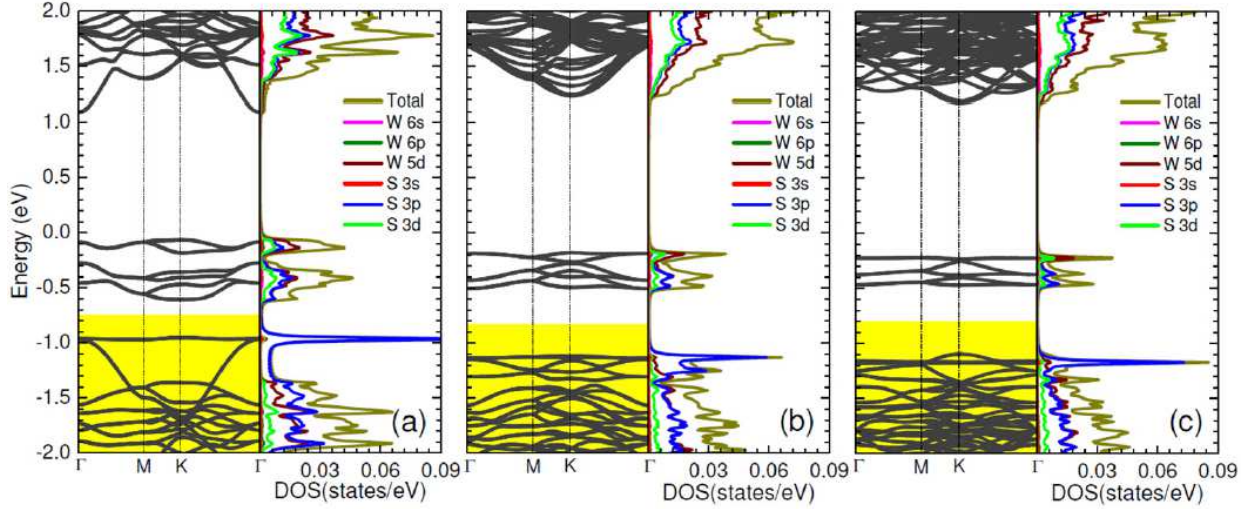}
\caption{The calculated band structure and corresponding partial
DOS of WS$_2$ monolayer with (a) 3$\times$3, (b) 4$\times$4, and
(c) 5$\times$5 supercells each containing a single W vacancy. The
intersection of white and yellow regions shows the Fermi energy.}
\end{figure}

Comparing Fig. 3 with Fig. 4, we find that the Fermi energy in
both MoS$_2$ and WS$_2$ monolayers is shifted down in energy by
the presence of metal vacancies. This suggests that the Mo/W point
defects can make the system a p-type semiconductor, in agreement
with DFT results \cite{Lu}. Moreover, the defect states of WS$_2$
are closer to VBM than that of MoS$_2$ monolayer, indicating that
the WS$_2$ monolayer may act as a more efficient p-type
semiconductor than MoS$_2$, when the metal vacancies are induced.
\begin{figure}
\center\includegraphics[width=1.0\linewidth]{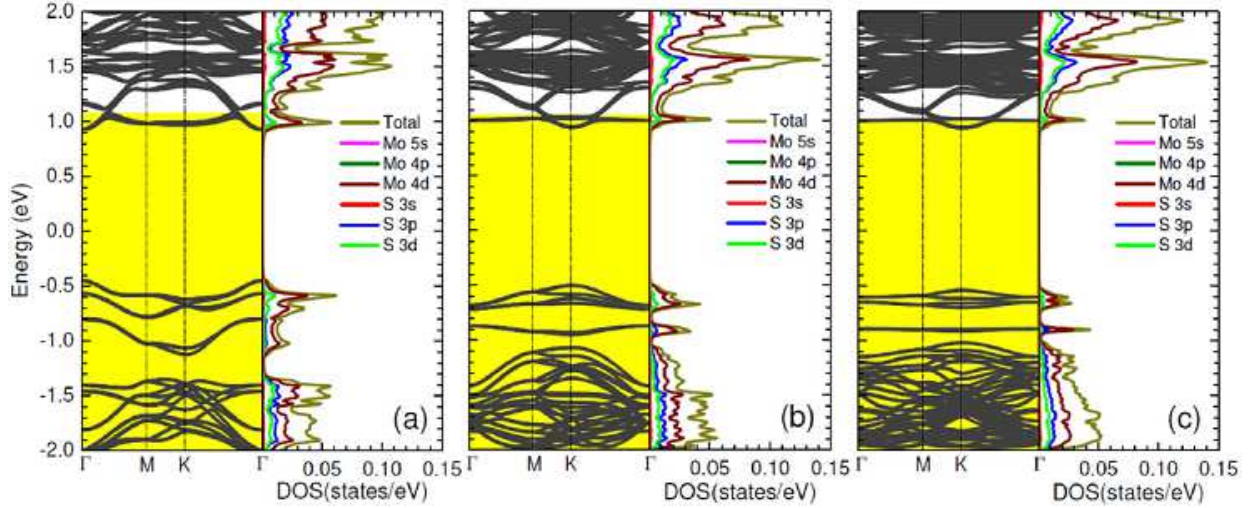}
\caption{The calculated band structure and corresponding partial
DOS of MoS$_2$ monolayer with (a) 3$\times$3, (b) 4$\times$4, and
(c) 5$\times$5 supercells each containing a single S vacancy. The
intersection of white and yellow regions shows the Fermi energy.}
\end{figure}

Let us study the influence of chalcogen defect on the electronic
structure of the monolayers. Figs. 5 and 6 show the band
structures and the partial DOS of MoS$_2$ and WS$_2$ monolayers,
respectively, when a single sulphur vacancy (Fig. 1(b)) is created
per supercell. From Fig. 5(a) it is evident that at high defect
concentration, the sulphur vacancies induce a midgap band with
bandwidth $\sim 0.6$ eV in the vicinity of the VBM, indicating
that the defect states tend to be more delocalized due to the
interaction between S vacancies. The midgap band manifests itself
as defect states with three peaks in partial DOS which become more
localized as the concentration decreases (Figs. 5(b) and 5(c)). In
addition, there is a flat band just below the MoS$_2$ CBM which
does not change notably with the concentration changes. The Mo
$4d$ orbitals around the vacancies play the main role in creation
of midgap states, while the S $3p$ and $3d$ orbitals do not
contribute considerably to the defect states of gap region. The
presence of sulphur vacancy shifts the Fermi level to the bottom
of the conduction band due to unsaturated electrons in the Mo
orbitals around the vacancy defect. This property suggests that
sulphur vacancies can turn the MoS$_2$ monolayers into a n-type
semiconductor in agreement with theory and experiment
\cite{Radisavljevic,Liu2,Lu,Qiu}.

On the other hand, the electronic band structure of WS$_2$
monolayer in the presence of sulphur vacancies with concentration
$\frac{1}{27}$ shows a narrow band with bandwidth $\sim 0.15$ eV
in the gap region (Fig. 6(a)). In this case the contribution of W
$5d$ orbitals and S $3p$ and $3d$ orbitals in generation of midgap
states are almost the same, as shown in the partial DOS of Fig.
6(a). As the size of the supercell increases, the defect state in
the bandgap becomes more localized around the vacancy region and
the contribution of W $5d$ states dominates in the localized state
as can be seen in the DOS of Figs. 6(b) and 6(c). The defect state
which acts as a donor level, lies about 0.6 eV below the CBM, in
agreement with the DFT result \cite{Wei}, indicating that at a
high enough temperature some of the localized charges can be
transferred to the CBM and therefore increase the system
conductivity. Accordingly, we conclude that the sulphur vacancies
act as electron donors and make the both MoS$_2$ and WS$_2$
monolayers electron rich. Besides the monosulphur vacancies, the
effect of two neighboring sulphur (disulphur) vacancies on the
same side of the monolayers, and also on different sides of the
monolayer, but on top of each other per supercell on the
electronic properties of the layers was also examined (not shown
here). Our TB calculations showed an additional shift of Fermi
level toward conduction band with an increase in the number of
localized states in the bandgap. The experimental observations
have demonstrated that these disulphur vacancies are less probable
to create, due to their formation energy which is roughly twice of
that of the monosulphur defect \cite{Zhou,Hong}.

\begin{figure}
\center\includegraphics[width=1.0\linewidth]{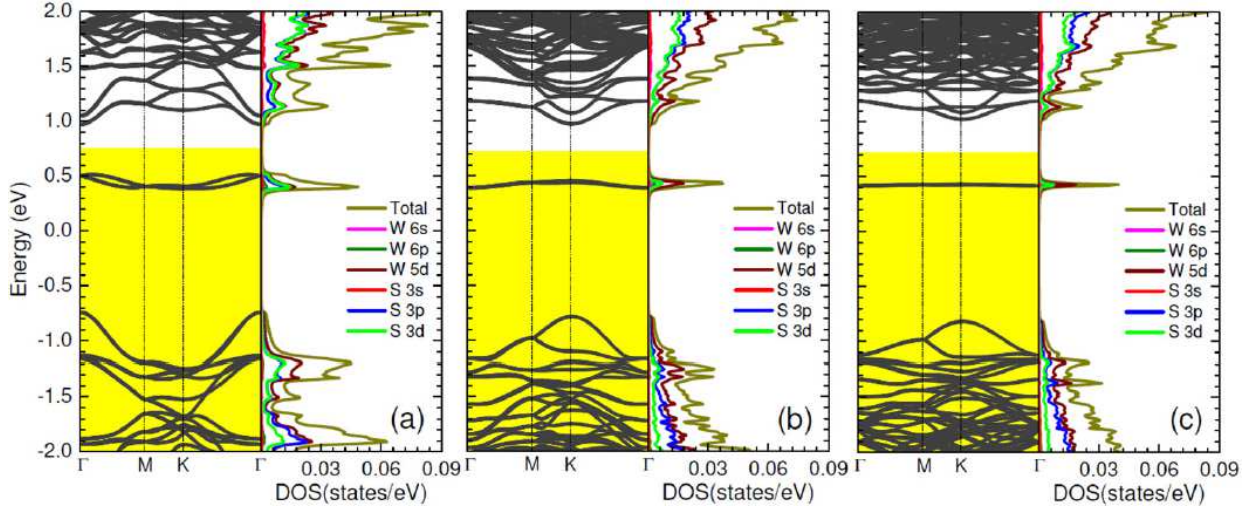}
\caption{The calculated band structure and corresponding partial
DOS of WS$_2$ monolayer with (a) 3$\times$3, (b) 4$\times$4, and
(c) 5$\times$5 supercells each containing a single S vacancy. The
intersection of white and yellow regions shows the Fermi energy.}
\end{figure}

Comparing the localized donor states and the Fermi energy in Figs.
5 and 6, it is clear that the n-type semiconducting behavior in
MoS$_2$ is more dominant than that in WS$_2$, when the sulphur
vacancies are introduced. This feature is in agreement with the
recent experimental observation of electronic properties of
MoS$_2$-WS$_2$ heterostructures, indicating that both MoS$_2$ and
WS$_2$ act as n-type semiconductors with relatively high Fermi
level in MoS$_2$ as compared to WS$_2$ \cite{Chen}.

To show the advantage of our tight-binding method over
first-principles calculations, we have also studied the vacancy
defects in the 11$\times$11 supercells which correspond to vacancy
concentration of $\sim 9.5\times10^{12}$ cm$^{-2}$. The electronic
structure of both MoS$_2$ and WS$_2$ monolayers with such a low
concentration of sulphur and metal defects are shown in Fig. 7.
Due to this low density of defects, the vacancy-vacancy
interaction is quite negligible and hence the midgap states are
strongly localized. Comparing Fig. 7 with Figs. 3-6, we see that
the p-type and n-type semiconducting behaviors in these defective
monolayers are not affected by the value of defect concentration.
Therefore, it is evident that this size of supercell is
computationally trivial for our tight-binding scheme, but
extremely expensive for DFT methods.

It is important to point out that the calculation of
spin-dependent density of states in the close proximity of sulphur
and metal defects did not show any spin polarization, indicating
that in the present approximation, the single S, Mo, and W
vacancies do not induce any magnetic moments.

Finally, we would like to emphasize that the agreement between the
results of our Slater-Koster tight-binding model and the
first-principles calculations in predicting p-type and n-type
semiconducting behaviors is related to the accuracy and
reliability of fitted parameters which provide us a more accurate
description of the band structures, as shown in Fig. 2.
Accordingly, our method is not only able to clearly demonstrate
the electronic band structure of defective MoS$_2$ and WS$_2$
monolayers, but also is very computationally affordable and can be
easily generalized to study very large systems with a random
distribution of single defects and other types of vacancies such
as MoS double vacancies, MoS$_2$ triple vacancies and antisite
defects \cite{Hong,Ataca}.
\begin{figure}
\center\includegraphics[width=0.8\linewidth]{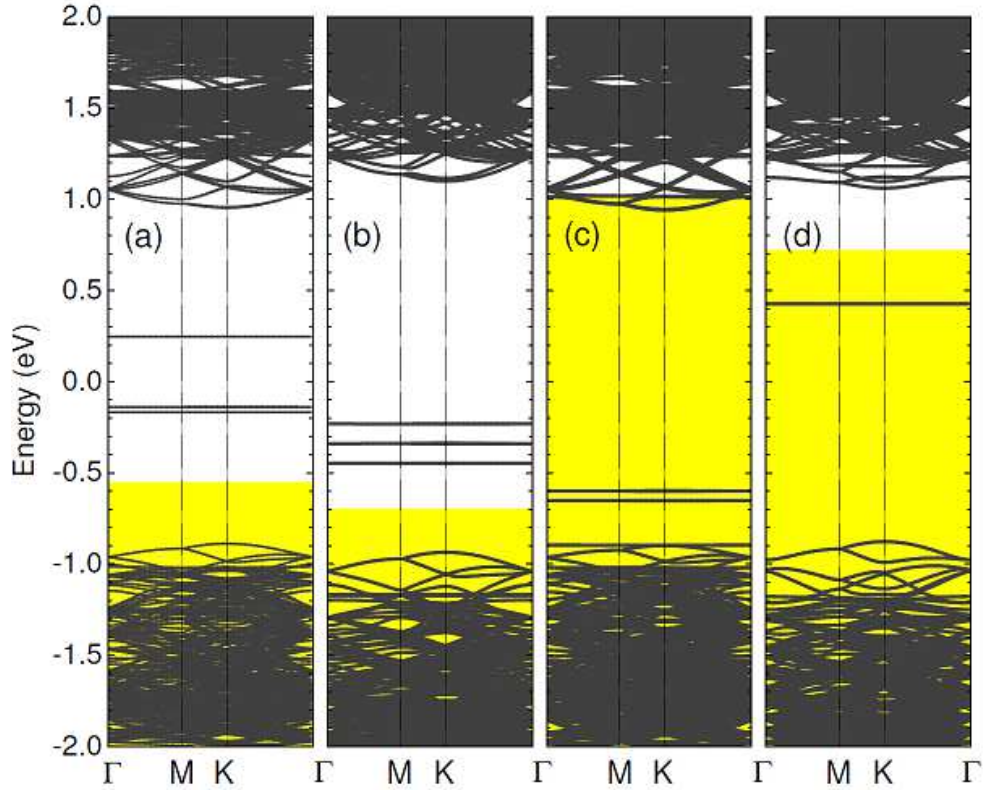}
\caption{The calculated band structure of MoS$_2$ and WS$_2$
monolayers with 11$\times$11 supercells each containing a single
atomic defect. (a) MoS$_2$ with Mo vacancies, (b) WS$_2$ with W
vacancies, (c) MoS$_2$ and (d) WS$_2$ with S vacancies. The
intersection of white and yellow regions shows the Fermi energy.}
\end{figure}

\section{conclusion}
Using Slater-Koster tight-binding model with non-orthogonal
$sp^3d^5$ orbitals and including the spin-orbit coupling, we have
explored the effect of atomic vacancies on electronic structure of
MoS$_2$ and WS$_2$ monolayers. The vacancy defects mainly create
localized states within the bandgap of pristine MoS$_2$ and
WS$_2$, along with a shift in the Fermi energy toward VBM or CBM.
As a result, the electronic properties of these monolayers are
strongly affected by the introduction of atomic defects. Our
results show that metal vacancies have potential to make the
monolayers p-type semiconductors, while the sulphur vacancies turn
the system as a n-type semiconductor.

Localization of midgap states by decreasing the defect
concentration in both metal and chalcogen vacancies suggests that
point defects in MoS$_2$ and WS$_2$ can act as resonant scatterers
\cite{Chen2,Saffar1}. Moreover, the vacancy-induced localized
states have the potential to activate new optical transitions with
energies less than energy gap in their optical spectrum,
suggesting a potential application of LTMDs for optoelectronic
devices.

\section*{Acknowledgement}
We are very grateful to F. Zahid and L. Liu for providing us the
WS$_2$ parameters and for valuable comments. This work is
partially supported by Iran Science Elites Federation.

\end{document}